# Effects of Hole-Transport Layer Homogeneity in Organic Solar Cells – A Multi-Length Scale Study

**Huei-Ting Chien*[1]**, Markus Pölzl[1], Georg Koller[2], Susanna Challinger[3], Callum Fairbairn[3] Iain Baikie[3], Markus Kratzer[4], Christian Teichert[4], Bettina Friedel[1]

[1]*Institute of Solid State Physics, Graz University of Technology, Austria;* [2]*Institute of Experimental Physics, Graz University, Austria;* [3]*KP Technology Ltd., Scotland, United Kingdom,* [4]*Institute of Physics, Montanuniversitaet Leoben, Austria*

**Abstract**

Irreproducibility is a serious issue in thin film organic photovoltaic (OPV) devices, as smallest local inhomogeneities can change the entire behaviour of identically built devices without showing obvious failure. Inhomogeneities can occur at various steps of device preparation and appear in all layers with different length scales and impact. The hole-transport interlayer (HTL) in OPV devices blocks unwanted electron diffusion to the anode and corrects energetic mismatch between oxide electrode and organic semiconductor. Most commonly used is commercial ink based on poly(3,4-ethylenedioxythiophene):poly(styrenesulfonate) (PEDOT:PSS) colloidal particles. However, exactly these are suspected to cause microscopic inhomogeneities, causing known irreproducibility of device characteristics. Considering PEDOT:PSS' acidity-caused electrode corrosion, it is questionable how much impact colloids have on device homogeneity. In this report, we give proof that a colloidal HTL does not necessarily cause device inhomogeneity and decreased efficiency, by comparing OPV devices with different HTLs, namely from commercial PEDOT:PSS ink and from $MoO_3$, obtained from two liquid precursors, leading to quasi-continuous or colloidal layers. With a combination of X-ray diffraction, atomic force and Kelvin probe microscopy, photoelectron and ambient air photoemission spectroscopy, we discuss the layers' properties from nano- to macroscale and demonstrate their impact upon implementation into OPV devices, via spatially-resolved characterization.

**1. Introduction**

Organic photovoltaic (OPV) cells have received considerable attention as a potential source of renewable energy due to their advantages of easy fabrication, light weight, low manufacturing cost and mechanical flexibility. Typically, an OPV device consists of a transparent conductive indium tin oxide (ITO) anode, a donor-acceptor bulk-heterojunction photoactive layer and a low-work-function metallic cathode[1-4]. To enhance the collection of photogenerated charges [5] and decrease leakage currents [6] a hole-transport layer (HTL) is commonly used between ITO and the active layer.

However, despite the fact that this type of devices reaches good efficiencies (presently best known 11% [7]), a major problem of organic thin film diodes is the significant batch-to-batch and device-to-device variation. Behaviour and performance of nominally identically built devices can differ considerably and so the research results published for identical systems. It is speculated that even device sets prepared at a time by the same experimentalist show not seldomly variations in performance of up to 10%.

Fact is that the physics of thin film diodes is naturally highly sensitive to smallest variations due to their limited lateral extension. These can be e.g. subtle differences in height, local composition, interface or contact properties, and the presence of impurities or grain boundaries. Such effects have been widely shown in the past for inorganic thin film solar cells.[8] In some cases the physical processes at such inhomogeneities were considerably altered from the expected device physics. Thereby the entire device may show a different behaviour to others of nominally identical built and not necessarily imply a clear device failure.

However, while organic devices are extensively studied on the nanoscale regarding their crucial donor-acceptor domain morphology, other rather microscopic variations originating from inhomogeneities of the electrode are often neglected. The few publications on this topic show its relevance, as the extensive study on organic solar cell degradation published by Hoppe et al.[9] and on microscale inhomogeneities in organic solar cells and modules been published by Revière et al.[10]. Both show that the effects are versatile, originating from surface roughness, work function fluctuations, activity/non-activity patches, grain boundaries, agglomerate formation, delamination, chemical reactions. And these rather intermediate-scale effects might not be visible via standard nanoprobing snapshots. In particular the poly(3,4-ethylenedioxythiophene):poly(styrenesulfonate) (PEDOT:PSS) electrode interlayer, widely used for its high conductivity, high transparency and solution processibility[11-12] is known to generate instability and shorter life times of devices especially at exposure to humidity, due to acidity and hygroscopicity of PSS [13-14]. Additionally, originating from synthesizing the insoluble PEDOT in presence of water-soluble PSS for better handling,[15] PEDOT:PSS is forming gel-like colloidal particles, which consist of a PEDOT-rich core covered by a PSS-rich shell,[16-18]. This colloidal form is suspected to be responsible for spatial inhomogeneities and irreproducibility [19].

Here, we suggest that the local microscopic inhomogeneities in solar cells caused by the PEDOT:PSS hole-conducting interlayer are independent of the colloidal film morphology. This is demonstrated by comparison of microscopic spatial characteristics in thin film OPV devices comprising HTLs of PEDOT:PSS and of sol-gel processed $MoO_3$ with either colloidal form comparable to PEDOT:PSS or a quasi-continuous featureless layer.

It has been demonstrated in the past that high work-function metal-oxides, such as NiO[20-21], $WO_3$ [22], $V_2O_5$ [23-25]and $MoO_3$ [5-6, 25-29] might be suitable alternatives for PEDOT:PSS, especially $MoO_3$, with its large bandgap between 2.9 to 3.1 eV and high work function 5.5 eV [25-26, 29-30]. In the present case, $MoO_3$ films are conveniently deposited from solution, equal to PEDOT:PSS. For that purpose sol-gel processing was used, allowing derivation of nano- to micron sized $MoO_3$ particle formulations from liquid precursors [5-6,27-28]. Thereby the morphology, electrical properties and surface physics of the HTLs are shown with X-ray diffraction, atomic force and Kelvin probe microscopy, photoelectron and ambient air photoemission spectroscopy, and their influence on OPV device physics demonstrated, by integrated device measurements and with spatially resolved photocurrent maps from nano- to macroscale. The architecture of the studied devices is shown in Fig.1. Herein, we use a blend of poly(3-hexylthiophene-2,5-diyl) (P3HT) and [6,6]-phenyl-$C_{61}$-butyric acid methyl ester ($PC_{60}BM$), a well-studied standard photoactive layer.

## 2. Experimental

### 2.1 Materials

Poly(3-hexylthiophene) (P3HT) was supplied by Rieke Metals Inc. ($M_W$ 50000-70000 g mol$^{-1}$, regioregularity 91-94%). [6,6]-Phenyl-$C_{61}$-butyric acid methyl ester ($PC_{60}BM$) was purchased from Nano-C Inc. (99.5% purity). The formulation of poly(3,4 ethylenedioxythiophene):poly(4-styrenesulfonate) (PEDOT: PSS) was purchased from Heraeus Deutschland GmbH & Co. KG (Clevios P Jet (OLED)). Ammonium molybdate ($NH_4)_2MoO_4$ (≥99.98%) was purchased from Sigma Aldrich. Hydrochloric acid HCl (≥37%) and bis(acetylacetonato)dioxomolybdenum(VI) ($MoO_2(acac)_2$) were bought from Sigma-Aldrich. Isopropanol was purchased from VWR International LLC. All the materials were used as received. ITO substrates (20 Ω/square, Ossila) were cleaned by sonication in acetone and isopropanol and followed by $O_2$-plasma etching (100 W for 30 min) briefly before use.

**2.2 Preparation of the $MoO_3$ and PEDOT:PSS HTLs**

Two different $MoO_3$ precursor formulations were used. The precursor leading to *continuous* films (***MoO₃-1***) was sythesized as follows: $MoO_3$ solution was prepared according to the procedure reported by K. Zilberberg et *al*. [27]. Here, $MoO_2(acac)_2$ was dissolved in isopropanol to form a 0.5% (w/v) solution. The precursor formulation resulting in *nanoparticle* films (***MoO₃-2***) was prepared by hydration method in aqueous solution as reported by Liu et *al.* [6]. Here, $(NH_4)_2MoO_4$ was dissolved in distilled water to form a 0.005 mol/L solution. Then 2 mol/L aqueous hydrochloric acid (HCl) was added drop-wise under stirring until the pH value of the solution was between 1 and 1.5. $MoO_3$ precursor formulations and PEDOT: PSS, were both filtered by 0.22 μm PVDF membrane filters (Sigma Aldrich) and spin-coated onto ITO substrates at 4000 rpm for PEDOT:PSS and 3000 rpm for $MoO_3$ for 40 sec, respectively. The layer of *MoO₃-1* was kept at ambient air for 1 hour for hydrolysis at room temperature and then annealed at 160°C for 20 min. The *MoO₃-2* film was directly annealed at 160°C in air (20 min). The PEDOT: PSS anode buffer layers were annealed at 160 ° C for 20 min under Argon (Ar) flow.

**2.3 Device fabrication**

Solar cells with P3HT:PCBM active layer according to the architecture in Fig.1 were prepared using different HTLs based on continuous $MoO_3$ (***MoO₃-1***), nanoparticle $MoO_3$ (***MoO₃-2***) and PEDOT:PSS, or were prepared without HTL (= *ITO*). For the devices, where applicable, hole-conduction layers were deposited on patterned ITO glass substrates and treated as described in 2.2. The active layer was applied in an in argon atmosphere by spin-coating from a solution of P3HT and PCBM (1:1 weight ratio, each 18 mg/mL) in 70°C chlorobenzene at 2500 rpm for 60 s, followed by annealing at 120° C for 10 min. The film thickness obtained is around 150 nm. The cathode was thermally evaporated as a bilayer of LiF (2nm)/ Al (100nm).

**2.4 Characterization**

Material structure and film morphologies were analyzed by atomic force microscopy (AFM) using two systems: For 1µm x 1µm a Nanosurf, EasyScan2, and for 3µm x 3µm and 100µm x 100µm a MFP 3D AFM System from Asylum Research. The images with the EasyScan2 were taken in tapping mode using a Tap190 cantilever (Budgetsensors, Bulgaria) with a nominal frequency of 190 kHz. The MFP 3D AFM was operated in intermittent contact mode using SSS NCHR AFM probes from APPNano with typical tip radii below 2 nm. Local contact potential difference (CPD) maps of 3µm× 3µm areas were measured by Kelvin probe force microscopy (KPFM) using the MFP3D system with Pt coated ACCESS EFM probes from APPNano. The AFM topography, according particle size distribution and contact potential difference (CPD) data were visualized and analyzed using the Gwyddion 2.40 software.

Surface CPD distribution maps of complete solar cell pixels (4mm x 1.5mm) area were examined by the SKP5050 Scanning Kelvin probe (KP Technology Ltd.). Work function results were obtained by Ambient Pressure Photoemission Spectroscopy System (APS) (APS02, KP Technology Ltd.) [31] under UV light source and ambient conditions with an excitation range of 3.3 - 6.8 eV. The thickness of the $MoO_3$ and PEDOT:PSS films was determined by variable angle spectroscopic ellipsometry (M-2000V, J.A. Woolam). The measurements were performed at three different angles (65°, 70°, and 75°) in the wavelength range of 200–1000 nm under three-layer optical model, silicon substrate, the native $SiO_2$ layer (1.7 nm), and the film bulk layer by the Cauchy function. The X-ray powder diffraction profiles were obtained on a Siemens D 501 diffractometer in Bragg-Brentano geometry using $Cu_{K\alpha}$ radiation (λ= 1.54178 Å) and a graphite monochromator at the secondary side. Data were fitted using EVA X-ray diffraction analysis software. X-ray photoelectron spectroscopy (XPS) measurements were performed on a commercial spectrometer (HR-XPS, SPECS Surface Nano Analysis GmbH), using monochromatic Mg $K_\alpha$ radiation (hν = 1253.6 eV). The illuminated current density–voltage (J–V) characteristics of the PV cells were measured using a computer-controlled Keithley 2636A source meter under AM1.5G illumination (100 mW/cm$^2$) from a solar simulator (Model 10500, ABET Technologies, rated ABB). Their external quantum efficiency (EQE) was recorded for wavelengths from 375 to 900 nm, using 250 W white light source (tungsten halogen) with monochromator, a computer-controlled Keithley 2636A source meter and a calibrated silicon photodiode. Spatial photocurrent distribution was scanned with a computer-controlled nano-manipulator-driven xy-stage (Kleindiek Nanotechnik, NanoControl NC-2-3) and excitation with a 532 nm laser (<5mW) with a focused spot-size of ≈2 μm.

3. Results

3.1 Properties of the hole-transport materials and interlayers

To study the pure colloidal effects of the hole-transport interlayer on the OPV device homogeneity and performance, commercial PEDOT:PSS was compared with two different types of hydrothermally grown $MoO_3$ particles, ***$MoO_3$-1*** as a continuous layer and ***$MoO_3$-2*** as a nanoparticle layer with a particle size comparable with PEDOT:PSS. XRD diffraction patterns of both $MoO_3$ films (see Fig.2) show a large number of diffraction peaks, the most prominent ones at 2θ = 9.7°, 19.5°, 25.8°, 29.4°, 35.5° and 45.5°, corresponding to the (100), (200), (210), (300), (310) and (410) crystal planes of the hexagonal $MoO_3$ phase (JCPDS Card No.21-0569, h-$MoO_3$), which is predominantly present in low-temperature syntheses[32-33]. The intensity distribution is not completely in agreement with the expected powder pattern according to the reference and also varies between the two film types. It can be assumed that there might be a small degree of preferential orientation in the films. Only two peaks, at 2θ = 32.8° and 40.0°, merely occurring in the continuous film *$MoO_3$-1*, could not be assigned to this $MoO_3$ phase, nor identified as another or precursor residues. X-ray photoelectron spectra (XPS) measurements were performed to confirm that both $MoO_3$ films have comparable surface chemistry. Fig.3 shows detail scans of the *Mo 3d* core level region of both $MoO_3$ films showing the characteristic Mo $3d_{3/2}$ (high binding energy) and Mo $3d_{5/2}$ (low binding energy) doublet. The doublet could be fitted with two peaks, one centered at 232.7 eV for Mo $3d_{5/2}$ and the other at 235.8 eV for Mo $3d_{3/2}$, respectively. These binding energies are consistent with literature values of $Mo^{6+}$ oxidation state of $MoO_3$[34-36]. The thicknesses of *$MoO_3$-1* and *$MoO_3$-2* films is in both cases around 10 nm and for PEDOT:PSS films around 40 nm, as measured by ellipsometry (not shown). The absolute work function of the HTLs on ITO substrates and of bare ITO was determined by APS [31]. Fig.4 (a) shows the according square-roots of photoemission of the various hole-conduction layers and ITO. Their work functions have been estimated from the offset of photoemission and found to be around 5.0 eV for PEDOT:PSS, *MoO$_3$-1* and *MoO$_3$-2* and around 4.6 eV for ITO. These results indicate presence of equal potential steps for hole collection from P3HT's highest occupied molecular orbital (HOMO)  and transfer to the ITO anode for any of the three hole conductors, as depicted in the energy level diagram in Fig.4 (b), showing the APS-derived work functions and reported molecular orbital energies  of P3HT, PCBM and work function of Al [6]  . In the wavelength spectral region relevant for solar cells (>350 nm), the optical transmission of the three HTLs is quite comparable with values close to 100%, as shown in Fig.5. Film morphology of the three HTL materials was investigated by AFM.  Fig.6 shows AFM height images of ITO on glass and of PEDOT:PSS in comparison with the continuous and nanoparticle MoO$_3$ films deposited on polished Si substrates.  The ITO surface (Fig.6a) shows the typical multicrystalline structure with quite high roughness of RMS = 3.54 nm. The PEDOT:PSS film (Fig.6b) shows a colloidal structure of almost spherical particles and considerable amount of agglomerates thereof resulting in an apparently quite broad particle size distribution with a mean diameter of 30 nm (Fig.6e) and film roughness of RMS = 0.95 nm (note: supplier gives 25 nm average size). The *MoO$_3$-1* film (Fig.6c) is the least rough one with RMS = 0.27 nm and shows the smallest sized particles (resolution was not sufficient to determine the shape) with a mean diameter of 6 nm and the most narrow size distribution (Fig.6e). Layers deposited from these particles are microscopically smooth compared to the other two materials, therefore in the following referred to as quasi-continuous. However, the present dark spots in the film indicate pin holes which were caused already during deposition by spin coating by evaporation of the solvent. The *MoO$_3$-2* film (Fig.6d) shows spherical particles with an average diameter of 18 nm, therewith slightly smaller than for PEDOT:PSS, also due to absence of aggregation, but shows a broader distribution of the primary particle size (Fig.6e) compared to  *MoO$_3$-1,* which induced higher RMS = 0.36 nm. A summary of mean particle size of the materials and roughness of the according HTL on silicon and on ITO (as described further below) can be found in Table 1.

The appearance of these structures on a larger scale, when applied on the transparent ITO electrode can be seen from Fig.7 showing AFM topography images of bare ITO and the different HTL coatings on ITO substrates. Expectedly, bare ITO (Fig.7a) exhibits strong surface roughness with RMS of 3.5 nm, caused by its multicrystalline structure. This roughness is well-known to induce difficulties and inhomogeneities for further deposition of the organic active layer. PEDOT:PSS is commonly used for its properties not only as HTL but also for establishing a more flat topography on rough ITO. In present case, with 40 nm of PEDOT:PSS deposited on ITO, the structure appears softened by the particles, leading to a reduced RMS of 1.1 nm, but the original crystal pattern is still visible (Fig.7b). In the case of $MoO_3$ HTLs, the film thickness must be much less than for PEDOT:PSS to avoid decreased photocurrent by optical absorption losses[27]. Therefore, the according films have no significant padding effect on the ITO structure, as can be seen for *$MoO_3$-1* (Fig.7c) and *$MoO_3$-2* films on ITO substrate (Fig.7d). In both cases, the rough ITO surface pattern is apparent, the $MoO_3$ hardly noticeable at this magnification. Thereby the roughness of ITO/*$MoO_3$-1* is with RMS of 1.4 nm still larger than for ITO/*$MoO_3$-2* is with RMS of 1.3 nm, probably because the continuous film rather replicates the underlying surface, while particles manage to fill "valleys". Potential work-function fluctuations depending on a colloidal or continuous HTL structure across the device area on the nano- and microscale might have effects on the spatially resolved and integral device function. This was investigated on ITO and various ITO/HTL configurations on different length scales and (where applicable) compared to according spatially resolved photocurrent maps of respective P3HT:PCBM solar cells with this HTL configuration. The nanoscale contact potential difference distribution between AFM probe and surface ($V_{CPD} = (\Phi_{Sample} - \Phi_{Tip})/e$) has been measured by KPFM across 3 µm x 3 µm areas, alongside with the aforementioned AFM topography images. The maps in Fig.7 e-h show the relative changes in contact potential across the surface (for better visibility ITO scaled separately, HTLs scaled to the maximum change among samples). All samples show spatial inhomogeneities of surface potential, expectedly the strongest fluctuations are shown by the bare ITO film (Fig.7e), with feature sizes correlating with the topographic features (Fig.7a). The large contrast of the pseudo color image, compared to any of the HTL configurations, also indicates quite high amplitude of these spatial fluctuations and found to be 50 mV, determined from maximum CPD peak-to-peak difference, which is consistent with previous KPFM measurements performed on ITO/Glass [37]. The samples with HTLs also show distinct features in the CPD maps but with considerably lower amplitude, being 20 mV for ITO/PEDOT:PSS, 22 mV for ITO/*MoO$_3$-1* and 15 mV for ITO/*MoO$_3$-2*. In the case of ITO/PEDOT: PSS (Fig.7f) the CPD map shows a pattern of lighter and darker regions, which is similar, but not entirely corresponding to the associated topography (Fig.7b), suggesting that there are additional variations caused by non-uniform surface potential of the PEDOT:PSS particle agglomerates themselves. The continuous MoO$_3$ film sample ITO/*MoO$_3$-1* (Fig.7g), does not reflect any features from the underlying ITO topographic structure (Fig.7c) in the CPD map, but shows a subtle fine pattern of slightly different potential regions and additionally some peculiar dark spots, i.e. localized circular areas of low potential. The latter might arise from aforementioned pin holes in the MoO$_3$ thin film, which were not even visible in the associated topography image. The fact that these pin holes appear to have a larger dimension on ITO than in the topography image on Si (Fig.6b) is plausible, because the granular surface structure of ITO works as local seed point for solvent evaporation. The difference between the pin holes' potential (on ITO potential) and the MoO$_3$ film surface causes a nominally higher CPD difference, than the film would have otherwise. The CPD map of the MoO$_3$ nanoparticle film sample ITO/*MoO$_3$-2* (Fig.7h) shows clearly the lowest amplitude of fluctuations in surface potential and the subtle brighter and darker areas in the film barely reflect the underlying ITO pattern (Fig.7d). From the results it can be seen that the amplitude of spatial potential fluctuations is generally reduced if any HTL is deposited, indicating that the lacking uniformity of the underlying ITO anode can be thereby greatly improved.  Further it can be concluded that the spatial variations of CPD seem to be independent of the padding effect an HTL material has on the underlying rough ITO structure. However, the surface coverage, as thin as it might be, seems to play an important role, as can be seen from the effect of pin holes for ITO/*MoO$_3$-1*. The fact that ITO/PEDOT:PSS shows the highest amplitude of surface potential fluctuations, despite the fact that it shows neither significant pores or reflects morphological features of underlying ITO surface or colloidal agglomerates, leads to the suspicion that the PEDOT:PSS material itself is non-uniform regarding its surface properties. This has been suggested in the past and based on observed conductivity anisotropy, PSS segregation or agglomerate formation [17-18, 38-39] . While photoconductive AFM allows visualization of nanoscale photocurrent distribution, this would be clearly dominated by features of donor/acceptor domains in the photoactive layer blend and not allow any further conclusions on the HTL effects [40-41] . A sub-microscale comparison of HTL surface potential with the photocurrent distribution, as derived by laser-beam-induced current mapping, was conducted. Fig.8 shows relative photocurrent density (*J*) maps of P3HT:PCBM solar cells with ITO, PEDOT:PSS, *MoO$_3$-1* and *MoO$_3$-2* with a scanning area of 100 μm x 100 μm and for comparison a CPD map inset of equal magnification with a scanning area of 30 μm x 30 μm. For the device with only ITO (Fig.8a), it can be seen that the observed nanoscale inhomogeneity of the ITO surface potential propagates also into larger scale (see inset) and seems to have detrimental effects on the photocurrent, because the cell shows large regions (about half the size of the scan area) of lower current output than in the rest of the area, with a difference in current density amplitude between these regions of about 0.5%. When any HTL is applied in the solar cell, the surface potential fluctuations (insets) are considerably reduced compared to ITO-only, as seen before on small scale. The device with PEDOT: PSS as HTL (Fig.8b) exhibits larger continuous areas of high current output with scattered small regions of low output, whereas the current amplitude between these areas changes again by 0.5%. In comparison, the cells with continuous HTL *MoO$_3$-1* (Fig.8c) and nanoparticle HTL *MoO$_3$-2* (Fig.8d) show a very different pattern. Despite the fact that no significant changes in surface potential are visible on this scale (insets) for either of them, the photocurrent map shows a small scale pattern of strongly scattered regions of extremely high, medium and very low current output with changes in current amplitude of 1% across the small area. Hereby the fractions of high photocurrent output make up 40% of the area for *MoO$_3$-1 and* 60% for *MoO$_3$-2* cells, the very low output regions make up about 5% and <1% of the device area, respectively. The very localized regions of extremely low output for the *MoO$_3$-1* cell might be caused by the pin hole effect which has been mentioned earlier. All-in-all within the four conditions, the device with the HTL of *MoO$_3$-2* has the best photocurrent homogeneity and highest density of high-output regions, indicating that the particles of MoO$_3$ are well distributed across the ITO, shielding its inhomogeneities very effectively despite the small nominal thickness and promoting efficient charge transfer at the electrode interface. Which consequences these local effects have on the entire device area is finally investigated in terms of photocurrent distribution on the complete pixel area of 4.0 mm x 1.5 mm size of a P3HT:PCBM solar cell and compared with equally large area CPD (measured by scanning Kelvin probe) of the associated ITO/HTL configurations, shown in Fig.9. The bare ITO electrode (Fig.9a), shows a very strong gradient in surface potential across the entire area with a maximum difference of 65.5 mV in a similar range, compared to the values from small area. For the photocurrent distribution this inhomogeneity causes equally strong fluctuations, whereas a large area of strong output is found in the center of the solar cell pixel and clearly decreasing outwards. Thereby the photocurrent density ($J$) amplitude changes by 12% across the pixel area. When PEDOT: PSS is used as HTL (Fig.9b), the fluctuations in surface potential across the area get much more refined and fluctuations less intense with a maximum difference of only 47.4 mV. In consequence, the photocurrent distribution of the according solar cells is much more uniform, showing a large almost homogenous area of reasonable but not extremely high output, with few negligible pixel edge effects, with changes in current amplitude of only 5%. However, there are clearly no high output areas on the PEDOT:PSS cell. The $MoO_3$-HTLs samples draw a different picture. ITO/*$MoO_3$-1* (Fig.9c) and ITO/*$MoO_3$-2* (Fig.9d) show both a similarly scattered pattern in their surface potential distribution, comparable to ITO/PEDOT:PSS. Thereby *$MoO_3$-1* shows still some areas of higher uniformity. The maximum fluctuations however, are with of 55.5 mV for *$MoO_3$-1* and 48.5 mV for *$MoO_3$-2*, also quite similar to PEDOT:PSS. This is a different trend than recorded on the small length scale. But though the surface potential distribution of the three HTL configurations on this length scale is similar, the output pattern of the solar cells with $MoO_3$ HTL are very different to the one with PEDOT:PSS. The photocurrent of the cell with *$MoO_3$-1* shows one large homogeneous area of high to very high output, with photocurrent amplitude fluctuation of only 2%. Thereby the minor defects (spots of low output), probably caused by aforementioned pin holes, were neglected. The *$MoO_3$-2* cell, shows equal behavior, but without defects, showing one large homogeneous area of high to very high current output, again with a maximum change in amplitude of 2%.

## 3.2 Device performance of solar cells

The integrated solar cell characteristics of P3HT:PCBM devices with the different electrode/HTL configurations ITO, ITO/PEDOT:PSS, ITO/*MoO$_3$-1* and ITO/*MoO$_3$-2*, have been determined by standard methods, to confirm the findings from the area sensitive characterization. Fig.10a shows a semi-logarithmic plot of the dark current density of the four systems. The device with only ITO shows clearly the largest leakage current, roughly one order of magnitude higher than the device with PEDOT:PSS. The two MoO$_3$ systems are located in-between, with about half an order of magnitude lower leakage current than for the ITO-only device. At higher forward bias, all four devices show an identical character, which can be expected as this part is dominated by bulk charge transport in the active layer, which is the same for all devices. Also the spectral response of the device is mostly determined by the active layer, as visible from the identical shape of the external quantum efficiency (EQE) of the devices shown in Fig.10b. Its amplitude however, differs slightly between the four cells, with the highest value for ITO/PEDOT:PSS with 57% EQE and the lowest for ITO with 51% EQE. As the EQE is determined by the short-circuit current ($J_{SC}$) per wavelength, its trend correlates directly with that of the $J_{SC}$ seen in the photocurrent characteristic recorded at 550 nm (close to the wavelength of maximum EQE) at same light intensity, as shown in Fig.10c.  Also the open-circuit voltages ($V_{OC}$) of the four cells vary, with lowest value for the ITO device with 0.440 V, highest $V_{OC}$ for ITO/PEDOT:PSS with 0.479 V, and the values of ITO/*MoO$_3$-1* and ITO/*MoO$_3$-2* equal with 0.467 V in-between. This poor $V_{OC}$ of the ITO device can be easily explained by voltage losses at the shunts, which were confirmed by the dark current characteristics. An additional observation is the lower rectangularity of the photocurrent curve of ITO/PEDOT:PSS compared to those of ITO/*MoO$_3$-1* and ITO/*MoO$_3$-2*, as visible from the lower fill factor (FF) of 0.57 for ITO/PEDOT:PSS, compared to 0.68 for ITO/*MoO$_3$-1* and 0.69 for ITO/*MoO$_3$-2*. This behavior can indicate interfacial barriers for charge transfer at the electrode interface or bad transport, which lead to accumulation of charges. The fact that this is already prominent at low light intensities, as they were used for EQE and monochromatic photocurrent measurement ($P_{monochr}$~3.5 mW/cm$^2$), strongly suggests that worse performance of the device can be expected at high light intensities, when a higher density of charges is created in the device. Photocurrent characteristics of the devices at high light intensities have been recorded under simulated solar conditions, i.e. white light AM1.5G illumination with P=100 mW/cm$^2$, as presented in Fig.10d. A summary of characteristic solar cell values is given in Table 2. The ITO device still shows the lowest $J_{SC}$ and $V_{OC}$ compared to the other systems, with 9.1 mA/cm$^2$ and 0.56V, respectively. Main reason for that is the unfavorable potential barrier between ITO and P3HT (Fig.4b). With 9.7 mA/cm$^2$, the ITO/PEDOT:PSS device exhibits a considerably lower $J_{SC}$ than ITO/*MoO3-1* with 10.3 mA/cm$^2$ and ITO/*MoO$_3$-2* with 10.5 mA/cm$^2$, while their $V_{OC}$ is identical with 0.59 V. As predicted, ITO/PEDOT:PSS exhibits an even more decreased FF at this light intensity of 0.46, similar to that of the ITO device. Both MoO$_3$ systems show higher FF of 0.50 for *MoO$_3$-1* and 0.55 for *MoO$_3$-2*. The lower value for the cell with continuous MoO$_3$-1 HTL could be caused by charge transfer issues in regions with pin holes. Altogether, this leads to the maximum power conversion efficiency *η* for the ITO/*MoO$_3$-2* device with 3.38%, followed by ITO/*MoO$_3$-1* with 3.05%, 2.64% for the ITO/PEDOT:PSS and 2.32% for the ITO device.

## 4. Discussion

It is obvious that the PV performance of a conventional organic solar cell is generally greatly improved by presence of a HTL, indicated by the fact that lower work function of ITO and direct contact between ITO and the organic semiconductor layer P3HT: PCBM cause charge blocking at the interface and large leakage currents within the cell. Comparing the colloidal PEDOT:PSS as HTL with two different $MoO_3$-HTLs of identical physical and morphological material properties, except for their form, continuous vs. nanoparticle colloidal layer, interesting observations: Despite the fact that a continuous layer with pin holes and a thin colloidal layer with potential voids both should show some sort of "porosity" allowing effects from the underlying ITO to shine through to the surface, the *$MoO_3$-2* nanoparticle HTL samples show most homogeneous surface potential and photocurrent distribution on small length scales, highest and most homogeneous photocurrent on the full device area, best integral device performance. The continuous *$MoO_3$-1* HTL configuration delivered also homogeneous potential and output, except for the regions with pin holes, visualized as localized spots of potential drops and low photocurrent output and in consequence slightly lower overall performance. PEDOT:PSS on the other hand, the long-term favorite among OPV HTLs, shows inhomogeneities in surface potential even on a very small scale, despite the fact that the thicker layer (of 40 nm) is efficiently padding the rough ITO surface (RMS 3.5 nm → 1.1 nm), indicating the variations arising from the material properties itself, e.g. aggregation, degradation (with In migration) or excess PSS segregation at the film surface[17-18]. In a solar cell, this effect causes patchy performance fluctuations on a small length scale, which seem to develop into charge transfer barriers on the large scale, as reflected in homogeneous but considerably lower output of the cells and finally low device efficiency at standard AM1.5G conditions. In summary, well distributed small particles of MoO$_3$ as HTLs in organic solar cells lead to a better spatially uniform photocurrent distribution and best PV cell electrical performance with an overall efficiency $\eta$ reaching 3.38 %. From the presented results it can be assumed that the two MoO$_3$ HTLs would lead to entirely identical performance in absence of the pin-holes. Therefore it is suggested that the failure of PEDOT:PSS device homogeneity and performance cannot be deducted from its colloidal state or surface coverage, but rather its chemical properties, such as PSS segregation or acidity-caused electrode corrosion.

## 5. Conclusions

We compared the surface and device inhomogeneity of P3HT: PCBM bulk heterojunction PV cells influenced by three different solution-processed colloidal HTLs, one PEDOT:PSS, one continuous MoO$_3$ and one nanoparticle MoO$_3$ film, and compared them with ITO only devices. This was supported by the comparison of the morphology and contact potential difference distribution of HTL layers and spatial photocurrent distribution of the OPV devices at different resolutions from the nano- to the micrometer scale.  This has been discussed in relation to the difference in integral device characteristics and performance between those different OPV cells. The results showed anode film homogeneity and device performance greatly improved by presence of any HTL. Regardless of continuous or nanoparticle layers, MoO$_3$ HTLs lead to entirely identical performance excluding the pin-hole effect, which induced slightly lower performance and uniformity. In contrast to the MoO$_3$ HTLs, PEDOT:PSS HTL showed spatial inhomogeneities and device charge transfer barriers, which may be caused by its chemical chracteristics. Independent of colloidal or continuous form, MoO$_3$ proves as a better candidate for anode buffer layers, leading to higher performance, higher homogeneity, and also lower cost, in solution-processed organic solar cells.

**Acknowledgement**

H.-T. C. and B.F. are grateful to the Austrian Science Fund (FWF) for financial support (Project No. P 26066) and to Dr Anna Coclite (TU Graz) for providing the ellipsometry measurements.

**Figures:**

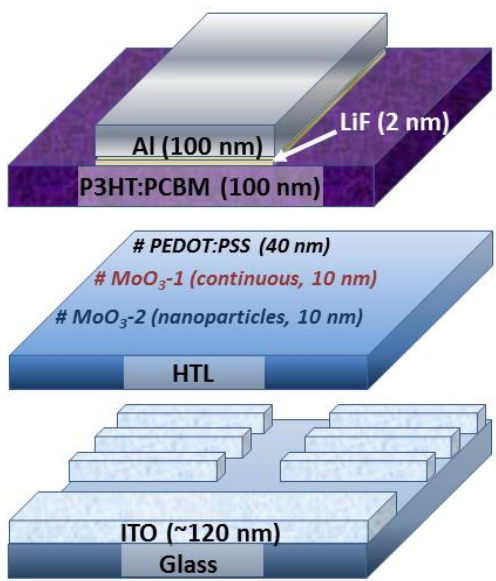

Fig.1 Device architecture of the studied solar cells, with according layer thicknesses noted.

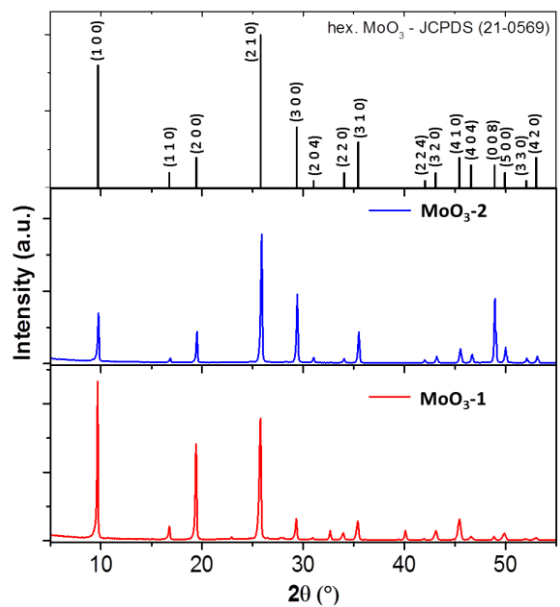

Fig.2 X-ray diffraction patterns of solution-processed continuous *MoO$_3$-1* (red line) and nanoparticle (blue line) *MoO$_3$-2* films in comparison with the reference pattern of hexagonal MoO$_3$ (black bars, JCPDS 21-0569).

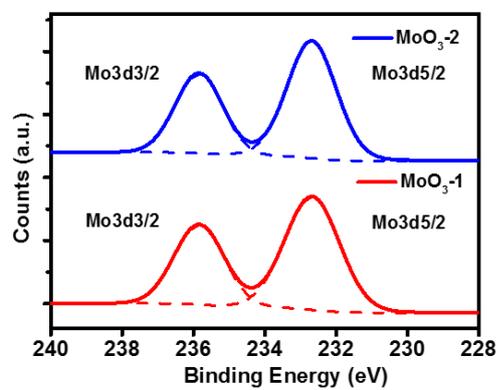

Fig.3 XPS spectra of different solution processed continuous *MoO$_3$-1* and nanoparticle *MoO$_3$-2* film, showing the Mo 3d core level spectra with the Mo 3d 3/2 and Mo 3d 5/2 peak doublet.

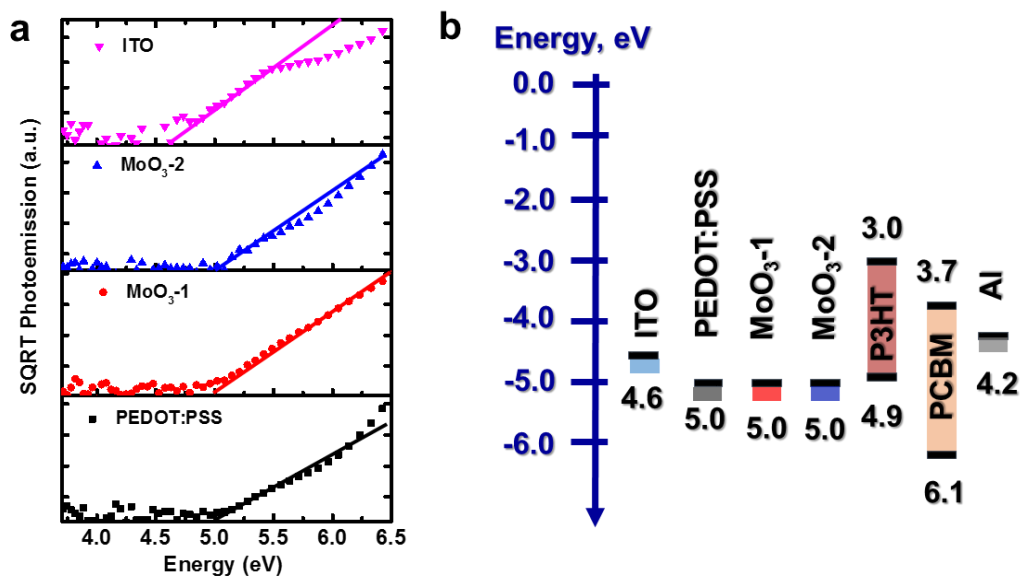

Fig.4 Intensity corrected normalized square-root of the air photoemission response of ITO and the different HTL films (a). Schematic energy level diagram with the accordingly derived work functions, and energies of additional components of the OPV structure, such as the HOMO/LUMO energies of P3HT and PCBM and the work function of Al from literature [6] (b).

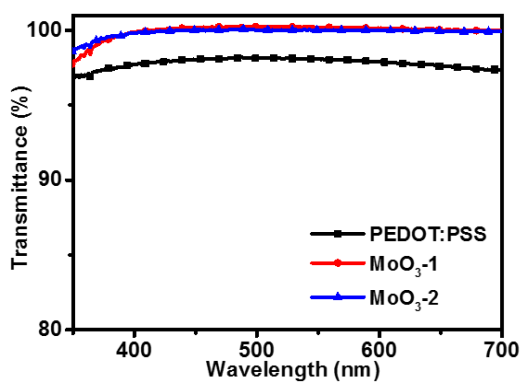

Fig.5 UV-Vis transmission spectra of different HTLs

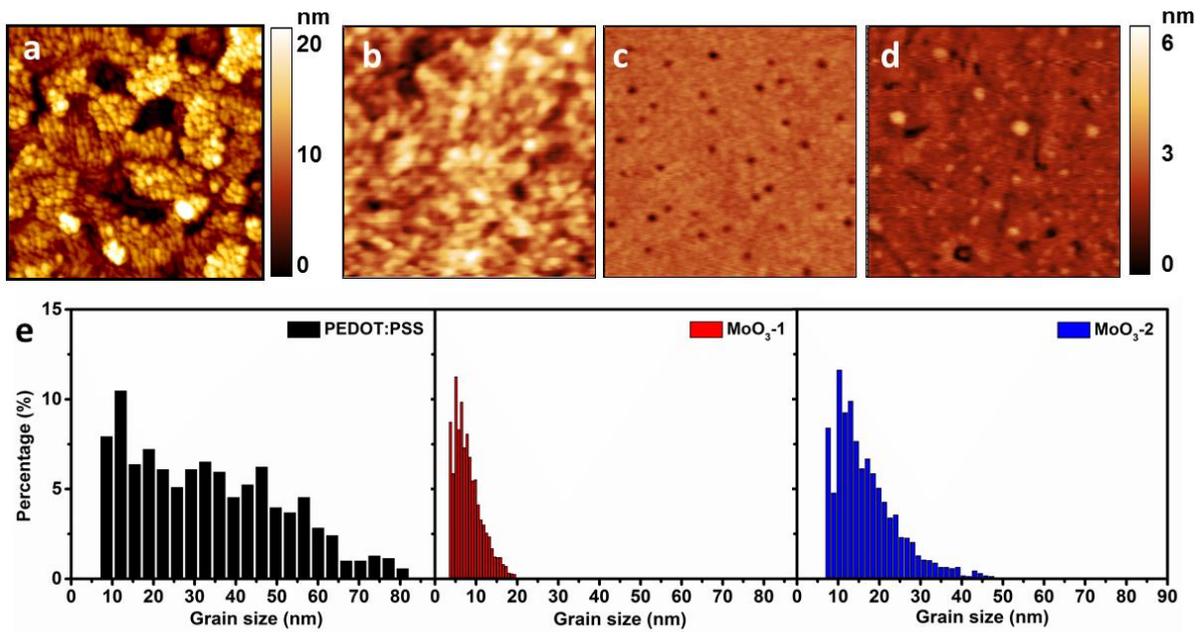

Fig.6 AFM topography images (1 μm x 1 μm area) of ITO on glass (a) and different HTL films on Si wafer substrates: PEDOT:PSS (b), *MoO$_3$-1* (continuous) (c) and *MoO$_3$-2* (nanoparticles) (d). Size distribution for PEDOT:PSS, *MoO$_3$-1* and *MoO$_3$-2*, as derived from AFM image particle analysis (e).

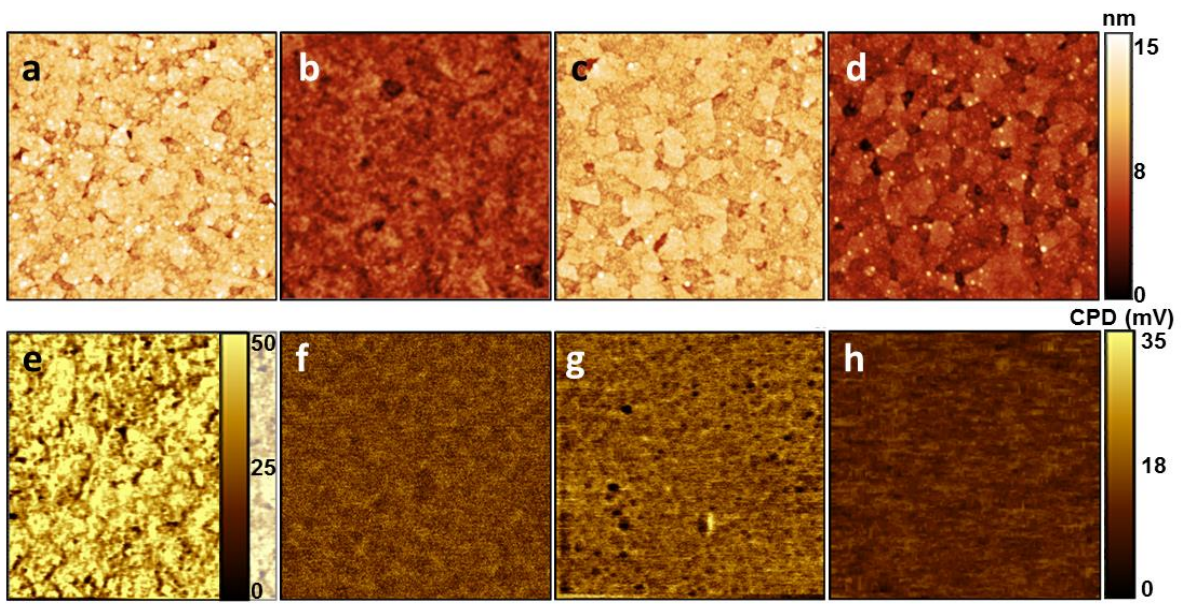

Fig.7 3 μm x 3 μm area images. AFM topography images (upper row) and the respective

relative surface potential maps (bottom row) of ITO (a+e), ITO/PEDOT:PSS (b+f), ITO/MoO$_3$-1 (continuous) (c+g) and ITO/MoO$_3$-2 (nanoparticles) (d+h).

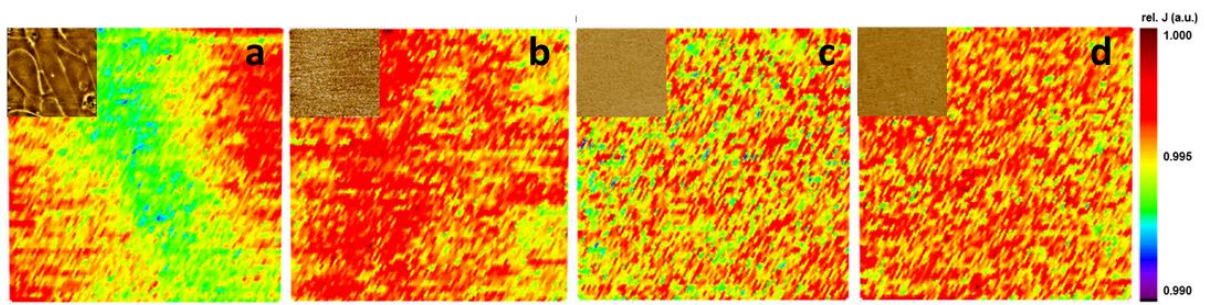

Fig.8 Relative photocurrent maps on the microscale (100 µm x 100 µm working area) of P3HT:PCBM solar cells without HTL (a) or with PEDOT:PSS (b), MoO$_3$-1 (continuous), (c) MoO$_3$-2 (nanoparticles) (d). Inset shows according underlying surface potential distribution map (30 µm x 30 µm).

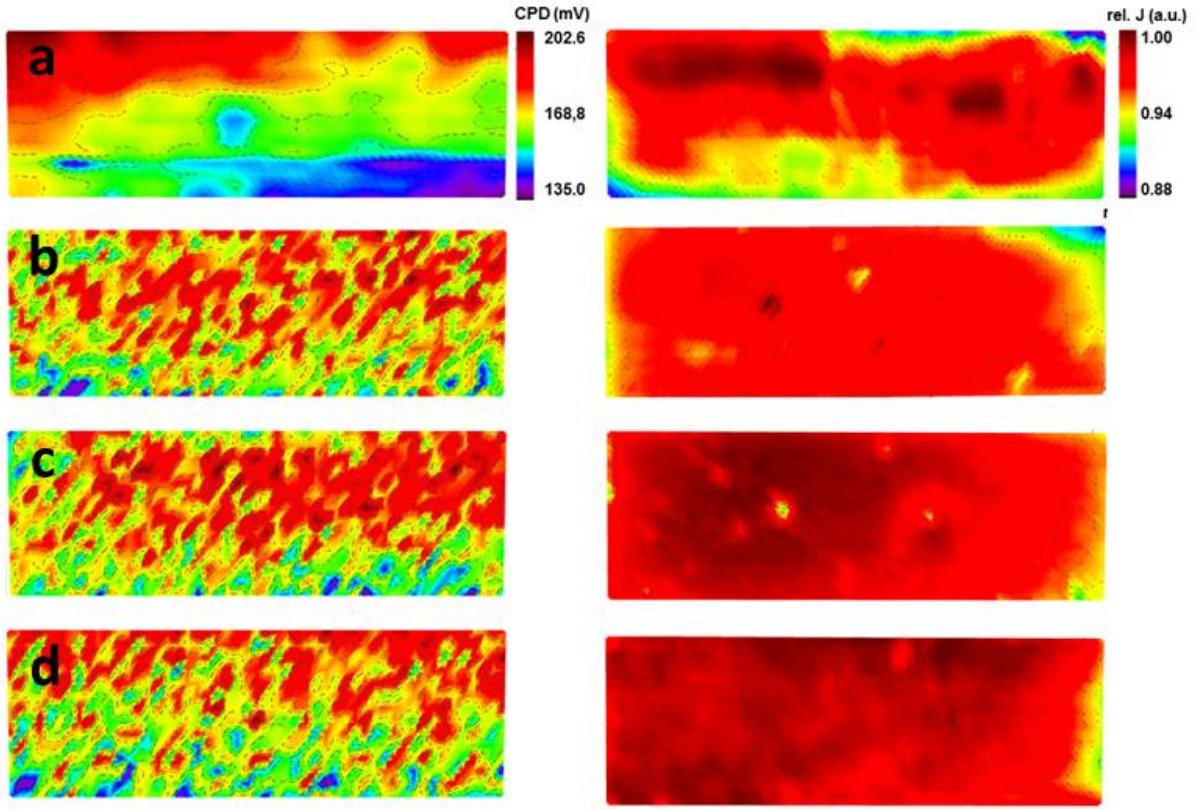

Fig.9 Large area surface potential (CPD) contrast of ITO/HTL films (left) and relative photocurrent map of according P3HT:PCBM solar cell pixel (right) for the configurations: bare ITO (a), ITO/PEDOT:PSS (b), ITO/MoO$_3$-1 (c) and ITO/MoO$_3$-2 (d). 4.0 mm x 1.5 mm scan area.

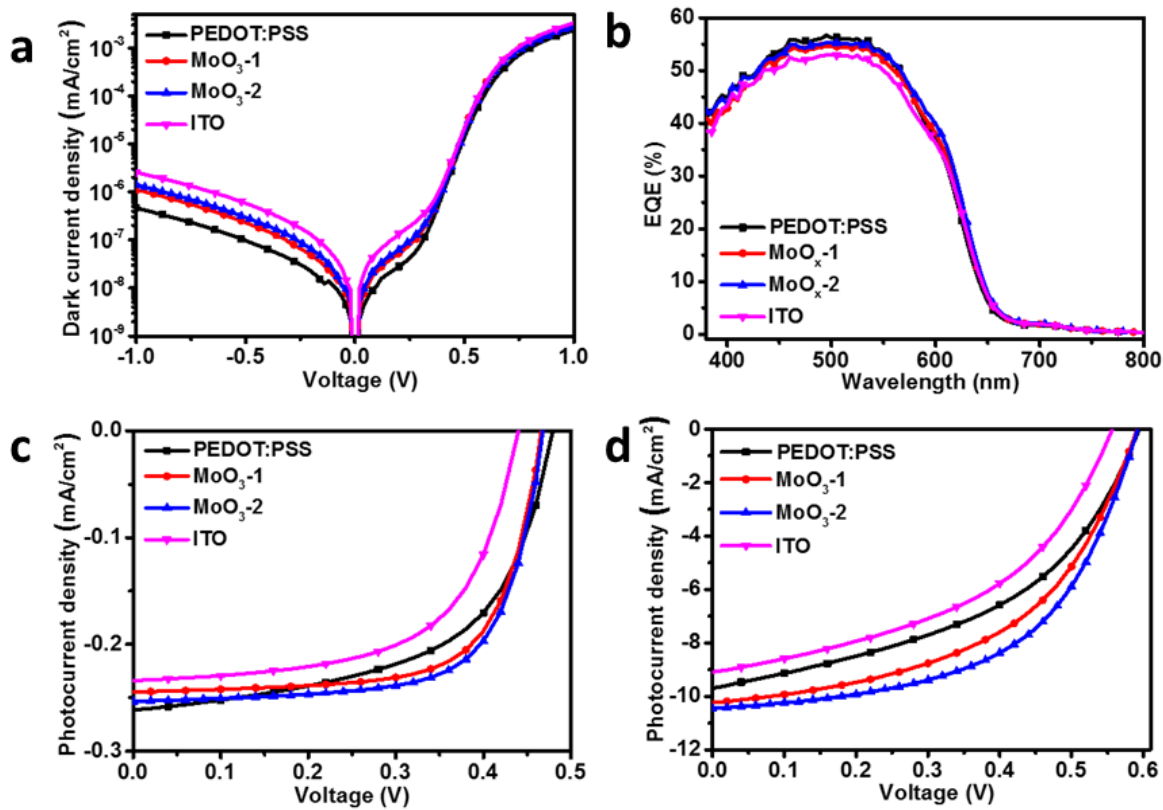

Fig.10 Electrical performances of the OPV devices with only ITO and with different HTLs. J–V characteristics in the dark (a). EQE (b) and J–V characteristics under monochromatic 550 nm illumination (c) at low light intensity of ~3.5 mW/cm$^2$. J–V characteristics under simulated solar illumination according to AM1.5G (d).

**Tables:**

Table 1 Mean particle size and film roughness of different HTL materials and films on Si and on ITO surface, as derived from AFM (Fig.6 and Fig.7 a-d).

| HTLs | Mean Diameter (nm) | $RMS_{Si}$ (nm) | $RMS_{ITO}$ (nm) |
|---|---|---|---|
| ITO | - | - | 3.5 |
| PEDOT:PSS | 30 | 1.0 | 1.1 |
| $MoO_3$-1 | 6 | 0.3 | 1.4 |
| $MoO_3$-2 | 18 | 0.4 | 1.3 |

Table 2 Summary of photovoltaic parameters of OPV cells with only ITO and with different HTLs, as derived from characteristics in Fig. 9d.

| HTLs | η (%) | FF | $V_{OC}$ (V) | $J_{SC}$ (mA/cm²) |
|---|---|---|---|---|
| ITO | 2.32 | 0.46 | 0.560 | 9.10 |
| PEDOT:PSS | 2.64 | 0.46 | 0.594 | 9.70 |
| $MoO_3$-1 | 3.05 | 0.50 | 0.591 | 10.24 |
| $MoO_3$-2 | 3.38 | 0.55 | 0.592 | 10.45 |